\documentclass[10pt,twocolumn]{article}  

\usepackage{geometry}
\usepackage{amsmath}
\usepackage{amssymb}
\usepackage{mathrsfs}
\usepackage{mathtools}
\usepackage[colorlinks=true,urlcolor=blue]{hyperref}

\usepackage{tikz}
\usetikzlibrary{calc,patterns,decorations.pathmorphing,decorations.markings,decorations.pathmorphing,angles,quotes,arrows,arrows.meta}
\tikzset{snake it/.style={decorate, decoration=snake}}

\usepackage{xcolor}
\definecolor{darkgreen}{rgb}{0,0.55,0}
\definecolor{midgreen}{rgb}{0,0.8,0.2}
\definecolor{magenta}{rgb}{1,0,1}
\definecolor{purple}{rgb}{0.5,0,0.5}
\definecolor{darkorange}{rgb}{1,0.55,0}
\definecolor{maroon}{rgb}{0.5,0,0}
\definecolor{olive}{rgb}{0.5,0.5,0}
\definecolor{midgrey}{rgb}{0.5,0.5,0.5}
\definecolor{lightgrey}{rgb}{0.75,0.75,0.75}
\definecolor{matlabblue}{rgb}{0,0.447,0.741}
\definecolor{matlabred}{rgb}{0.85,0.325,0.098}
\definecolor{lightblue}{rgb}{0,0.5,1}
\definecolor{darkgrey}{rgb}{0.25,0.25,0.25}
\definecolor{teal}{rgb}{0,0.5,0.5}
\definecolor{navy}{rgb}{0,0,0.5}
\definecolor{goldenrod}{rgb}{0.85,0.6,0.1}

\newlength{\mylinelength}
\newlength{\mydashlength}
\newlength{\mydashspace}
\newlength{\mylinethickness}
\newcommand{\myline}[1]{{\parbox{\mylinelength}{#1\rule{\mylinelength}{\mylinethickness}}}}
\newcommand{\mydash}[1]{{\parbox{\mylinelength}{#1\rule{\mydashlength}{\mylinethickness}\hspace{\mydashspace}\rule{\mydashlength}{\mylinethickness}}}}

\usepackage[numbers,sort&compress]{natbib}
\begin{document}


\title{Broadband control of water wave energy 
amplification in chirped arrays}

\author{A.~J.~Archer, 
H.~A.~Wolgamot, 
J.~Orszaghova 
\\
{\footnotesize{}
Faculty of Engineering, Computing and Mathematics, University of Western Australia, WA 6009, Australia}
\\
{\footnotesize{}
Wave Energy Research Centre, University of Western Australia, WA 6009, Australia}
\\[6pt]
L.~G.~Bennetts
\\
{\footnotesize{}School of Mathematical Sciences, University of Adelaide, SA 5005, Australia}
\\[6pt]
M.~A.~Peter
\\
{\footnotesize{}
Institute of Mathematics, University of Augsburg, 86135 Augsburg, Germany}
\\
{\footnotesize{}
Augsburg Centre for Innovative Technologies, University of Augsburg, 86135 Augsburg, Germany}
\\[6pt]
R.~V.~Craster
\\
{\footnotesize{}
Department of Mathematics, Imperial College London,
London SW7 2AZ, UK}
\\
{\footnotesize{}
Department of Mechanical Engineering, Imperial College London,
London SW7 2AZ, UK}
}%

\date{}

\twocolumn[
  \begin{@twocolumnfalse}
    \maketitle
    \begin{abstract}     
Water waves in natural environments are typically broadband, nonlinear and dynamic phenomena. Taking concepts developed for slow light in optics, we address the challenge of designing arrays to control the spatial distribution of wave energy, and amplify target frequencies at specified locations. Experiments on incident waves interacting with a chirped array of eight vertical cylinders 
demonstrate significant amplifications as predicted numerically, and provide motivation for application to energy harvesting.
\end{abstract}
  \end{@twocolumnfalse}
]


The ability of chirped (or graded) arrays to create slow light or sound waves, confine them, and filter them spatially according to frequency is attractive for wave control. This is often connected with rainbow trapping of light \cite{hess2007} or sound \cite{acou_rain,jimenez17a}, 
and, when combined with the optical properties of surface plasmon polaritons \cite{pendry2004}, 
has led to the development of surface dispersion engineering devices based on chirped gratings \cite{gan11a} or tapered devices \cite{argy13a}. 
A recurring theme is that by using a graded surface grating, different wavelengths are trapped at different locations. Improvements in design and fabrication of nanopatterned metal surfaces, 3D printing of surfaces and optimization of their optical and acoustic properties open new applications to exploit the effect in light storage, energy harvesting, filtering and delayed delivery for novel devices \cite{boriskina13a}. 
 
 This comprehensive body of work, primarily in optics, plasmonics and electromagnetism, forms a paradigm for the control of waves more broadly and has been extended to the field of elasticity \cite{celli1,krodel2015,colombi16a}, where it is used to generate mode conversion devices that hybridise surface waves to body waves. 
 A feature less often exploited is that energy can accumulate in regions where the group velocity tends to zero \cite{romero13a,sensing}. Assuming the local behaviour in the array is dominated by neighbouring elements enables the implied local periodicity to predict band-gaps (or forbidden frequencies), and the band-gap edges 
 determine a relation between local array properties and frequencies at which group velocities are zero.
 This interpretation 
 empowers design of broadband arrays with spatial filtering by frequency and localised amplification, 
 and has recently been taken into linear water wave theory \cite{bennetts_graded_2018}. 
 The approach complements other recent activity,
 in which ideas from transformation optics have been applied to water waves at the millimetric amplitude scale to achieve localised amplitude amplification up to a factor of three \cite{li18a}, 
 and constant-spaced arrays involving resonant components,  in the sense of metamaterials,  
 have been used to achieve low-frequency filtering with potential applications to coastal defence \cite{dupont_type_2017}.
 
\begin{figure}[b!]
    \centering
    \includegraphics[width=\columnwidth]{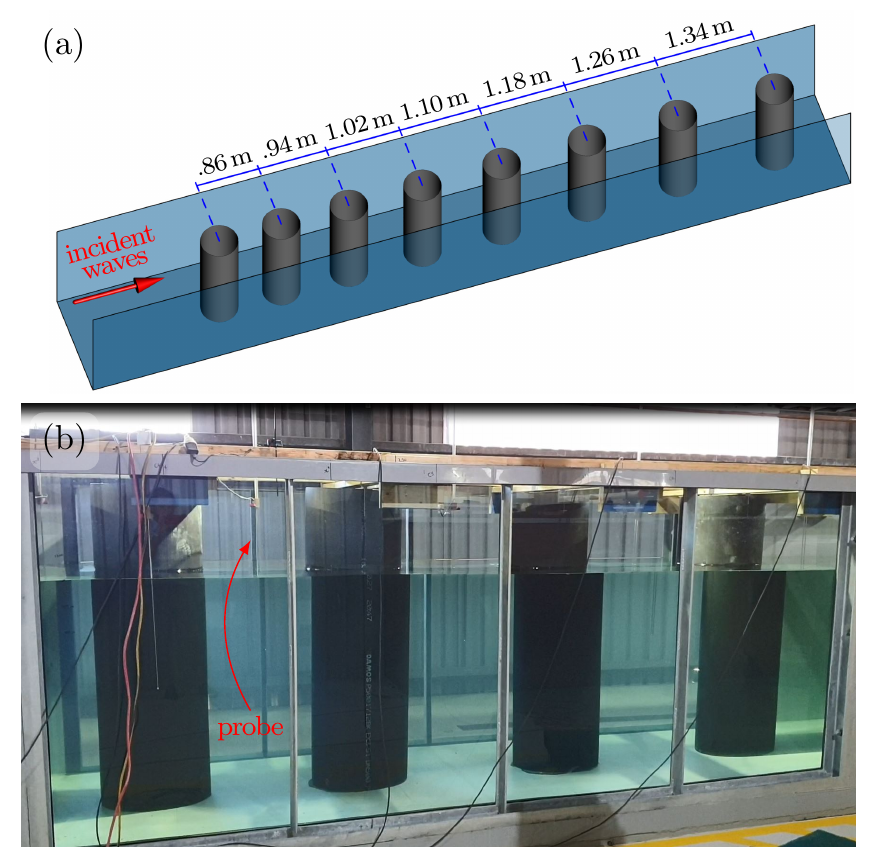}
    \caption{(a)~Wave flume experimental set-up, involving a chirped array of eight identical cylinders, with spacing increasing in the incident wave direction. 
    (b)~Side-view photo showing cylinders~3--6,
    including probes in cylinder spacings.}
    \label{fig:model}
\end{figure}

In a seemingly disparate branch of ocean engineering, 
there has been intense activity on methods to harvest the vast quantity of renewable energy carried by ocean waves. 
Many wave energy converter (WEC) designs have been proposed \cite{babarit2017}, 
and a very large class of WECs are tuned to resonate with the dominant frequencies in a sea-state,
but suffer from having small bandwidth compared to the bandwidth of the (random) incident waves \cite{falnes2007}. In any practical application, multiple interacting WECs will be present, and much research has focussed on regular arrangements of these \citep[e.g.][]{tokic19a}.

In this Letter we address the pressing question of whether it is possible to take advantage of generic concepts from rainbow trapping from optics in a (potentially nonlinear and dissipative) water wave system, 
to yield amplification of chosen frequencies at different locations, 
where (in an absorbing array) the energy could conceivably be captured by suitably tuned WECs. 
We do so through careful experimentation on waves of length order metres interacting with
an array of eight identical fixed bottom-mounted cylinders, of radius 0.25\,m  
arranged along the centreline of a 1.49\,m wide, 54\,m long, 1.6\,m tall wave flume, 
and with chirped spacing, similar to \cite{cebrecos_enhancement_2014}
but with the cylinder spacing increasing in the incident wave direction
(see Fig.\ \ref{fig:model}). 
The cylinders span the full height of the flume, which is filled with water to 1.1\,m depth.
A hinged wavemaker 
at one end of the flume 
generates incident waves, 
and those transmitted through the array are absorbed by a passive beach, which reflects $<$ 1\% of the energy reaching it (in the frequency range of interest).




\begin{figure*} 
    \centering
    \includegraphics[width=\textwidth]{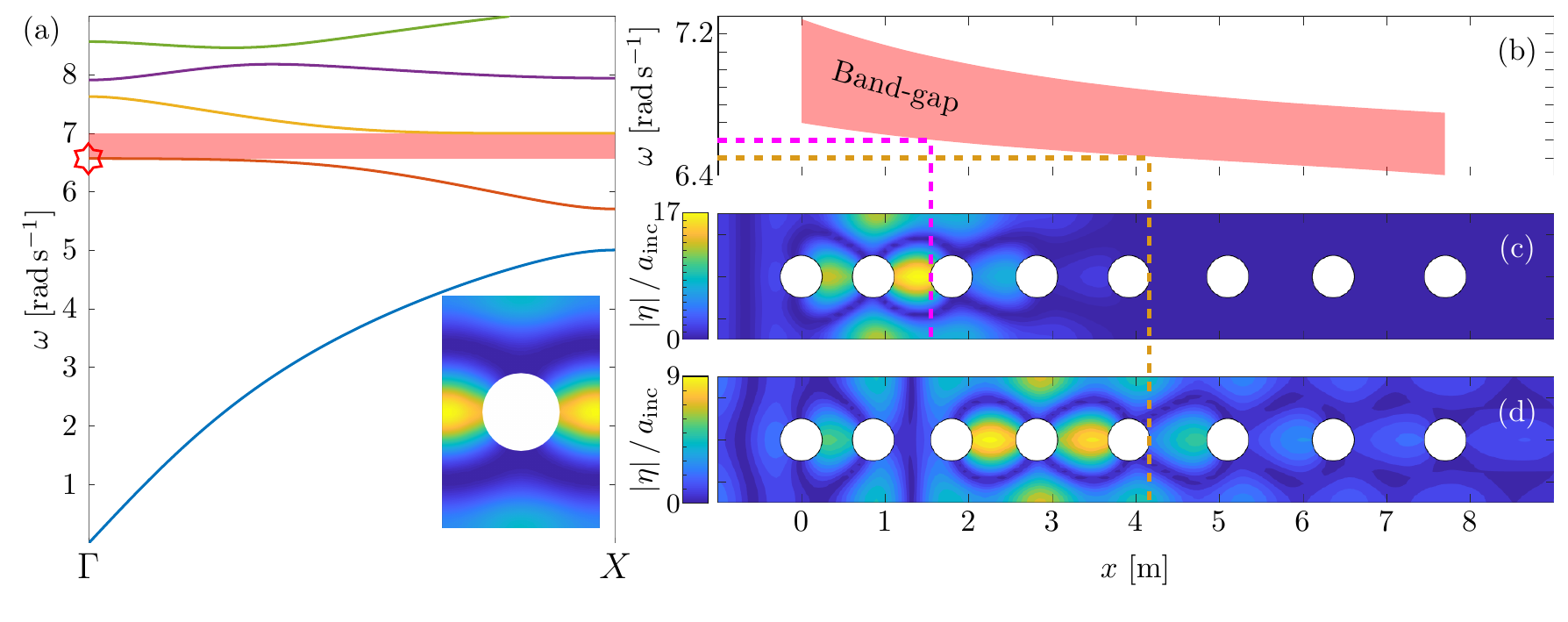}
    \caption{(a)~Band diagram (dispersion curves) of symmetric modes for infinite array, 
    using third cylinder spacing for $x$-periodicity,
    where the abscissa covers wavenumber space for normal incidence in the first irreducible Brillouin zone from in-phase ($\Gamma$) to out-of-phase ($X$), and red shading indicates second band-gap. Inset shows eigenmode on second band at $\Gamma$, as indicated by red star.
    (b)~Red shading shows evolution of the second band-gap as cylinder spacing increases along the array, where spacing is interpolated between spacing mid-points and extrapolated at ends.
    (c,d)~Model predictions of free surface elevation modulus, $\vert\eta\vert$,
    normalised by the incident amplitude, $a_\textnormal{inc}$,
    for regular incident waves with angular frequency (c)~$\omega=6.60$\,rad\,s$^{-1}$ and (d)~$\omega=6.50$\,rad\,s$^{-1}$.
    Dashed curves connecting (b) with (c,d) indicate theoretical predictions of zero group velocity, where $\omega=6.60$\,rad\,s$^{-1}$ (\mydash{\color{magenta}}) and $\omega=6.50$\,rad\,s$^{-1}$ (\mydash{\color{darkorange}}) enter the band-gap.}
    \label{fig:band}
\end{figure*}


Analysis of doubly-periodic infinite arrays---with periodicity in the propagation direction ($x$-direction) matching the local cylinder spacing, and the periodicity in the transverse direction matching the flume width (analogous to the walls)---yields valuable information to guide the experimental tests. 
The band diagram associated with the third spacing (Fig.\ \ref{fig:band}a) shows well defined band-gaps, with the
first two bands 
well separated from the more densely arranged bands above,  and informs choices of incident frequencies. We choose to operate towards the upper end of the second band as the wavelength is smaller than on the first band, so the waves scatter more strongly from the cylinders, travel more slowly and are in deep water relative to wavelength. The second band is very flat close to the upper band edge over a large range of wavenumbers, generating a high density of states, 
and enabling us to harness ``flat band slow light", thereby reducing dispersion \cite{li08a}.  

Simulations using the boundary element method \citep{sauter11a},
with channel walls accounted for using the decomposed tank Green's function \citep{chen1994side}, 
give the steady-state linear free surface responses around the array shown in Figs.\ \ref{fig:band}(c,d). 
The almost flat dispersion curve acts to slow the wave within the array, and the cylinder spacings at which the band-gap causes propagation to cease are predicted (see Fig.\ \ref{fig:band}b). 
The linear theory for an infinite array, implicitly assuming adiabatic grading of the array, is remarkably accurate in its predictions even for merely eight cylinders; moreover, the amplitude amplifications in the cylinder spacings before propagation ceases increase by an order of magnitude, i.e., factors 9 and 17 for $\omega=6.50$\,rad\,s$^{-1}$ 
(wavelength 1.46\,m; Fig.\ \ref{fig:band}d) 
and $\omega=6.60$\,rad\,s$^{-1}$ 
(wavelength 1.41\,m; Fig.\ \ref{fig:band}c), respectively.

\begin{figure*} 
    \centering
    \includegraphics[width=\textwidth]{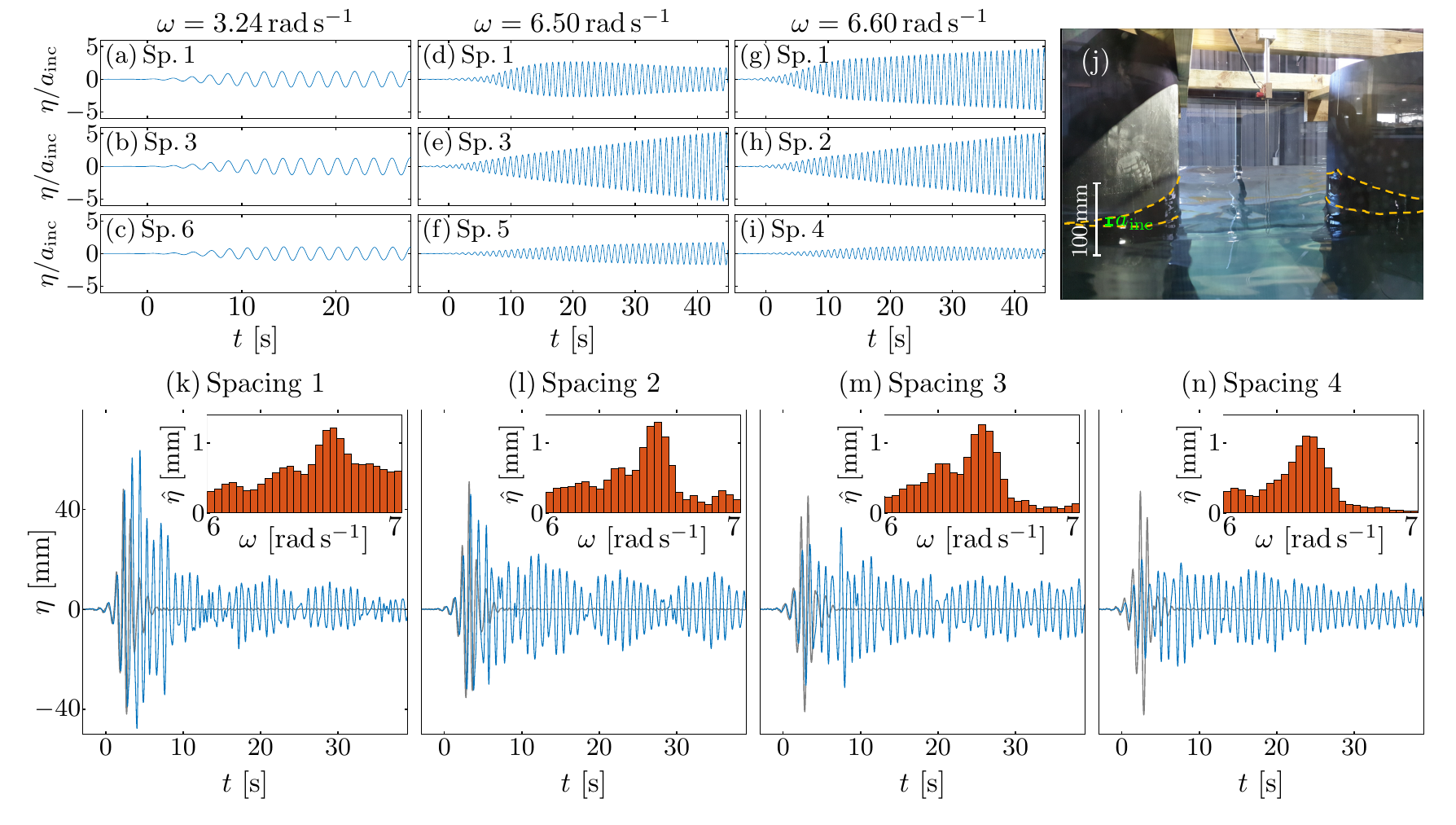}
    \caption{(a--i)~Linearised experimental time series for regular incident waves at frequency (a--c)~$\omega=3.24$\,rad\,s$^{-1}$, (d--f)~$\omega=6.50$\,rad\,s$^{-1}$, and (g--i)~$\omega=6.60$\,rad\,s$^{-1}$, 
    with responses shown in selected spacings,
    and where $t=0$ is the incident wave arrival time.
    (j)~Photo of resonant mode shape in spacing~3 for $\omega=6.50$\,rad\,s$^{-1}$.
    (k--n)~Experimental time series for focussed incident wave packet in spacings~1--4 (\myline{\color{matlabblue}}), 
    with corresponding incident packets superimposed (\myline{\color[rgb]{0.5,0.5,0.5}}).
    Insets show spectral decompositions of responses.
    }
    \label{fig:fig3}
\end{figure*}

 Given predictions of such large amplifications at precise localisations, 
 the wave flume experiments are used to explore the limitations of linear theory, and whether assumptions, such as ignoring viscosity, surface tension and wave breaking, 
 alter these predictions in practice. 
 We probe the resonant modes using regular incident waves of 
 angular frequency $\omega=6.50$\,rad\,s$^{-1}$ and 6.60\,rad\,s$^{-1}$ 
 (for comparison with theoretical responses in  
 Figs.\ \ref{fig:band}c,d), 
 and amplitude $a_\textnormal{inc}=0.01$\,m. 
 An incident wave of frequency $\omega=3.24$\,rad\,s$^{-1}$ and amplitude $a_\textnormal{inc}=0.03$\,m, 
 which exists on the first band and does not intersect a band edge, 
is also tested. 
 Eulerian point measurements of the free surface elevation
 are recorded with wave probes located at the centre of each of the first six spacings 
(Fig.\ \ref{fig:band} suggests the maximum response will be at the centre).
Band-pass filtered, normalised, linear time-series responses for
$\omega=6.50$\,rad\,s$^{-1}$ and 6.60\,rad\,s$^{-1}$ (Figs.\ \ref{fig:fig3}d--i)
show large amplifications in spacing~3 and spacing~2, respectively,
and minimal transmission farther along the array,
consistent with theoretical predictions.
Further, similar mode shapes to those predicted by linear theory are indicated by the wetted surface of the cylinders (Fig.\ \ref{fig:fig3}j). 
In contrast, the incident wave at frequency $\omega=3.24$\,rad\,s$^{-1}$  propagates through the array almost unchanged (Figs.\ \ref{fig:fig3}a-c). 

The time series in Figs.\ \ref{fig:fig3}(a--i) are truncated just prior to the point
when waves reflected by the array then re-reflected by the wavemaker return to contaminate the results.
The resonances have timescales longer than the test window available 
 \cite[see Supplementary Material,][]{Supp},
so amplitudes are still increasing towards steady-state at the end of the window,
and the maximum possible amplifications are not achieved. 
Numerical simulations do not suffer from contamination by spurious reflections, 
and suggest substantial increases in amplitude from 50\,s to 100\,s (35\% for $\omega=6.50$\,rad\,s$^{-1}$
and 55\% for $\omega=6.60$\,rad\,s$^{-1}$), 
and comparable increases from then till the steady-states are reached for $t>300$\,s
\cite[see][]{Supp}, 
although it is likely that 
viscous dissipation would result in smaller increases in the experiments if the test windows could be enlongated, even if wave breaking could be avoided.  

Focussed incident wave packets are used 
to investigate the overall spectral structure of the response.
The propagating linear free surface signal created at the wavemaker is 
\begin{equation}
    \eta_\textnormal{inc}(\tau)=\frac{A\,\sum_n \left\{S(\omega_n)\,\cos(k_n\,x_0-\omega_n\,\tau)\right\}}{
    \sum_n S(\omega_n)},
\end{equation}
where $A=0.045$\,m is the nominal amplitude, $x_0=-24$\,m is the mean wavemaker position relative to the linear focal point, $\tau$ is time relative to linear focus time, $k_n$ is the wavenumber, 
and $S$ is a Gaussian with  
standard deviation $\sigma=0.15$\,rad\,s$^{-1}$ and mean $\omega_\textnormal{p}=6.03$\,rad\,s$^{-1}$, i.e.\ a frequency on the second band.  The linear focus position is 2\,m into the array and the sum is over $2^{10}$ equally spaced components.
Figs.\ \ref{fig:fig3}(k--n) show raw surface-elevation time-series responses in spacings~1--4, 
with corresponding incident series superimposed for comparison (from tests without cylinders).
The insets show the response amplitude spectra in 0.04\,rad\,s$^{-1}$ frequency bins, 
from the (discrete) Fourier transform of the surface elevation time series, $\hat{\eta}=\vert\mathcal{F}(\eta)\vert$.
The spectral peak occurs on the second band, 
and is frequency downshifted from
$6.66\pm0.02$\,rad\,s$^{-1}$
to $6.42\pm0.02$\,rad\,s$^{-1}$
over
spacings~1--4, 
consistent with the theoretical predictions (cf.~Fig.\ \ref{fig:band}b).

\begin{figure} [!ht]
    \centering
    \includegraphics[width=\columnwidth]{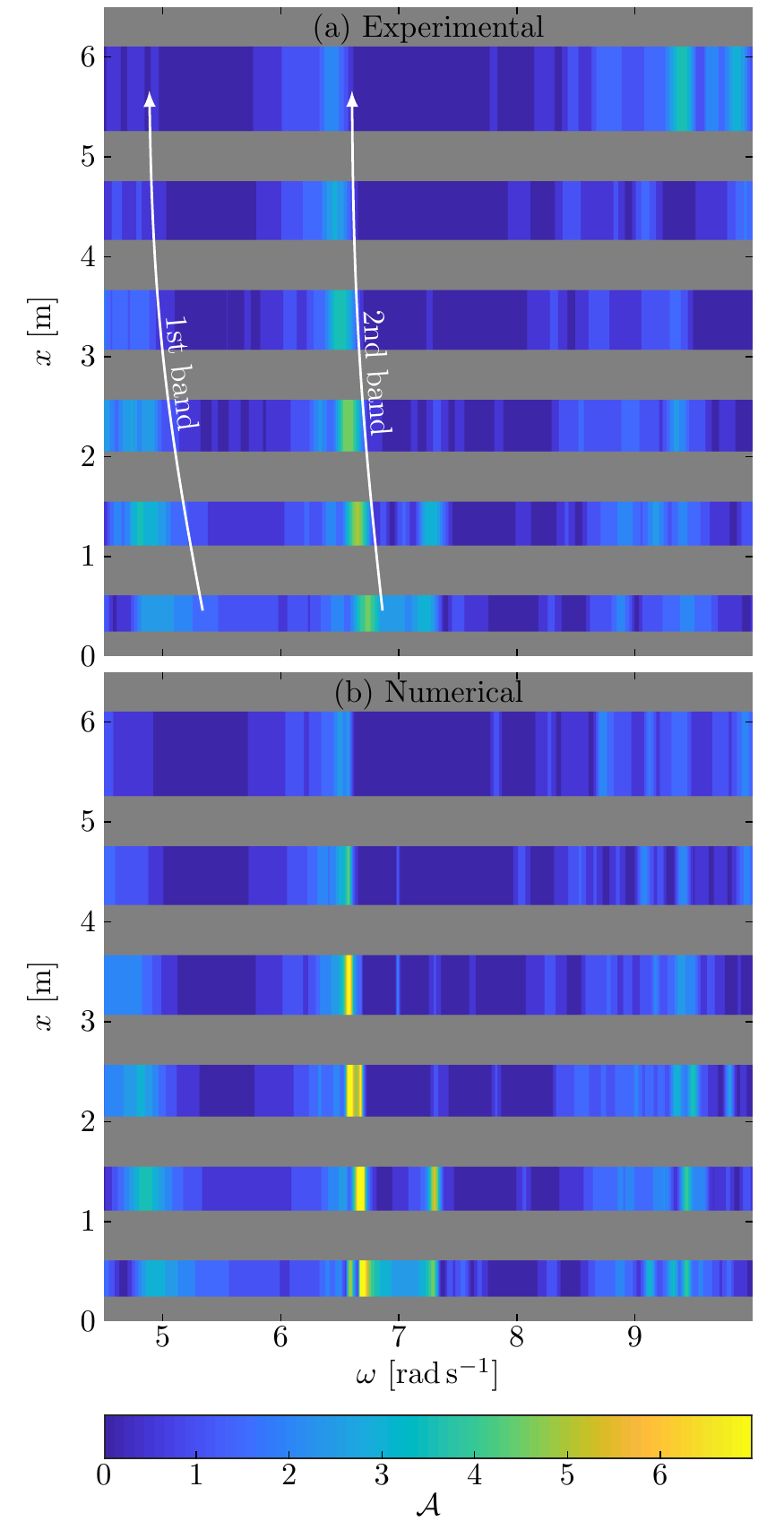}
    \caption{(a)~Experimental and (b)~numerical amplitude transfer functions, at wave probe locations along the array. 
    Grey strips represent spatial locations of cylinders~1--7.}
    \label{fig:figure4}
\end{figure}

A transfer function is defined as 
\begin{equation}
    \mathcal{A} = \frac{\hat{\eta}}{\hat{\eta}_\textnormal{inc}}
    \quad\textnormal{where}\quad
    \hat{\eta}_\textnormal{inc} = \vert\mathcal{F}(\eta_\textnormal{inc})\vert,
\end{equation}
to quantify amplitude amplifications over the frequency spectrum.
Fig.\ \ref{fig:figure4} shows the amplitude transfer function calculated from the numerical model (for which $\mathcal{A}=\vert\eta\vert\,/\,a_\textnormal{inc}$) and from 
focussed wave group experiments (averaged over  test 
shown in Figs.\ \ref{fig:fig3}k--n and tests with a shifted peak frequency 
$\omega_\textnormal{p}=6.55$\,rad\,s$^{-1}$ and shifted amplitude $A=0.029$\,m).
The experimental and numerical transfer functions are in excellent agreement, 
notwithstanding the expected larger and sharper amplifications in the 
numerical model.
The amplifications on the second band are evident,
along with the spectral downshift along the array.
Cognate amplifications are also shown for the first band  \citep[see][]{Supp}.

\begin{figure} [htb!]
    \centering
    \includegraphics [width=\columnwidth] {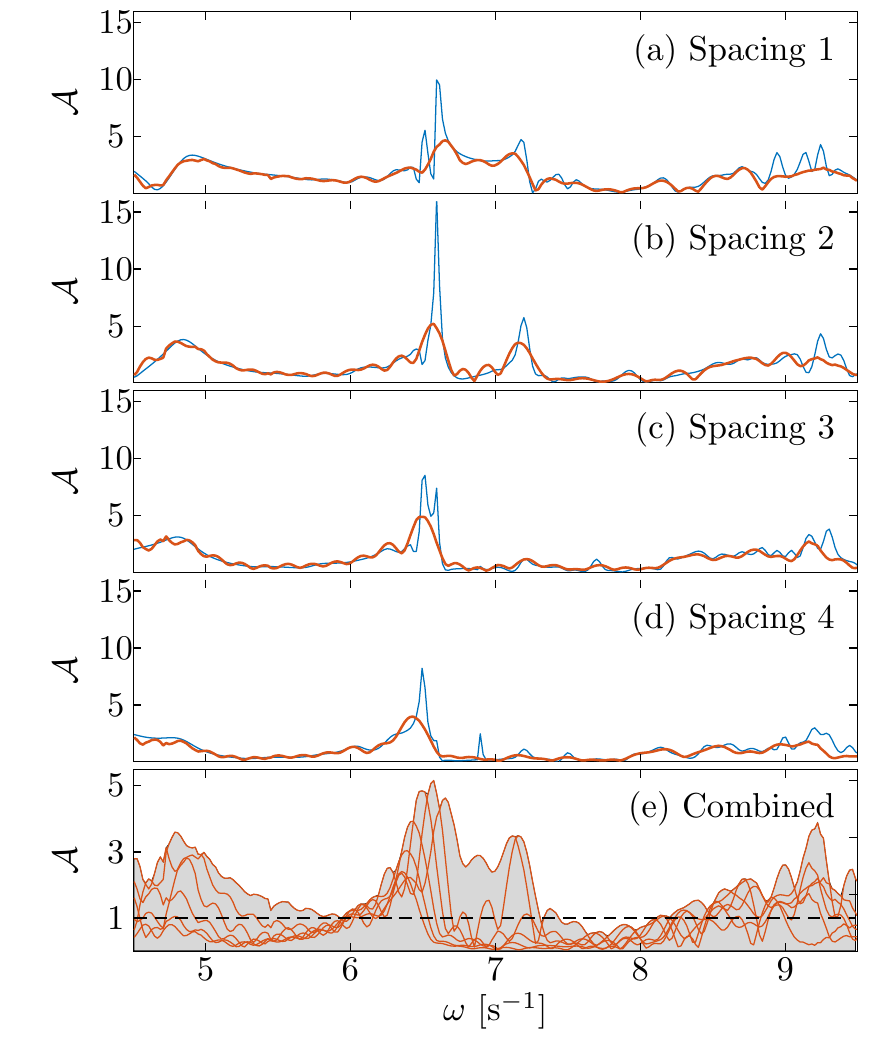}
    \caption{(a--d)~Experimental (\myline{\color[rgb]{0.85,0.325,0.098}}) and numerical (\myline{\color[rgb]{0,0.447,0.741}}) amplitude transfer functions in spacings~1--4.
    (e)~Combined experimental transfer functions for spacings~1--6
    (\myline{\color[rgb]{0.85,0.325,0.098}})
    with envelope (\protect\tikz \protect\fill [color=gray,opacity=0.5] (0,0)  rectangle (8pt,6pt);), and
    amplification boundary
    (\mydash{\color{black}}).}
    \label{fig:figure5}
\end{figure}

The numerical--experimental agreement is more clearly illustrated in Figs.\ \ref{fig:figure5}(a--d) for spacings~1--4.
The transfer functions overlap, except at the most resonant peaks,
where differences can partly be attributed to truncation of the experimental data resulting in under-estimation of the maximum amplifications. In the regular wave tests, 
the slow resonant build-ups are truncated (Figs.\ \ref{fig:fig3}e,h),
whereas in the focussed packet tests, 
truncations cut off the slow resonant decays.
Dissipative phenomena 
also contribute to the disparity, particularly at higher frequencies.

The broadband amplification given by the array is illustrated by Fig.\ \ref{fig:figure5}(e),
which shows the experimental transfer functions for spacings~1--6,
and with the envelope indicated by the grey shading.
The majority of the envelope lies well above the boundary $\mathcal{A}=1$, which divides amplification ($\mathcal{A}>1$) from suppression ($\mathcal{A}<1$),
with the only interval experiencing suppression from 7.45--8.10\,rad\,s$^{-1}$.
The maximum amplifications are $\approx{}5$, and occur on the second band, with corresponding energy amplifications $>20$.
Amplifications $>3$ also occur over multiple frequency intervals,
giving energy amplifications $\approx{}10$.


In this Letter we have presented results from wave-flume experiments 
that demonstrate localisation and amplification of water waves in a chirped array of bottom-mounted cylinders,
despite the challenges of working with highly resonant systems in a finite-length flume.
Linear theory was shown to predict the amplification spectrum 
accurately in frequency--distance space, 
including peak resonance locations and 
spectral downshifts along the array.
Moreover, even for the rather small array of eight cylinders used, 
the amplification locations, 
at which the group velocity slows to zero, 
were shown to be consistent with the band structure of infinite periodic arrays using local cylinder spacings.
The outcomes suggest future strategies for ocean wave energy harvesting should consider taking advantage of the broadband response and precise control demonstrated here.  
We have not attempted a parametric study to optimise such an array; 
this will be a future undertaking.

\section*{Acknowledgements}

AJA was supported by the Wave Energy Research Centre, jointly funded by The University of Western Australia and the Western Australian Government, via the Department of Primary Industries and Regional Development (DPIRD).  
HAW acknowledges financial support from Shell Australia. 
LGB is supported by an Australian Research Council mid-career fellowship (FT190100404).
RVC thanks the UK EPSRC for their support through Programme Grant EP/L024926/1 and also acknowledges the support of the Leverhulme Trust.

\end{document}